\documentclass[10pt,journal]{IEEEtran}
\usepackage{threeparttable}
\usepackage[T1]{fontenc}
\usepackage[numbers,sort&compress]{natbib} 

\usepackage{algorithm}
\usepackage{amsmath,amssymb,amsfonts}
\usepackage{algorithmic}
\usepackage{enumitem}
\usepackage[final]{graphicx}
\usepackage[caption=false,font=footnotesize]{subfig}
\usepackage{flafter}
\setkeys{Gin}{draft=false}
\usepackage[hidelinks]{hyperref}
\usepackage{textcomp}
\usepackage{xcolor}
\usepackage{booktabs}
\usepackage{pifont}
\usepackage{tabularx}
\usepackage{dsfont}
\usepackage{multirow}
\def\BibTeX{{\rm B\kern-.05em{\sc i\kern-.025em b}\kern-.08em
    T\kern-.1667em\lower.7ex\hbox{E}\kern-.125emX}}

\begin{document}

\title{Minimizing Bottleneck Overload Risk in Online Edge Task Placement under Workload Uncertainty}

\author{Lei~Wang,
        Winston~K.G.~Seah, 
        Gang~Xu$^{*}$,
        Xiaodong~Xu
        and~Baoqi~Huang%
\thanks{$^{*}$Corresponding author: Gang Xu (e-mail: csxugang@imu.edu.cn).}%
\thanks{L. Wang, G. Xu, X. Xu and B. Huang are with the College of Computer Science, Inner Mongolia University, Hohhot 010021, China.}%
\thanks{W. K. G. Seah is with the School of Engineering and Computer Science, Victoria University of Wellington, Wellington 6140, New Zealand.}%
}

\maketitle

\begin{abstract}
In edge computing, tasks must be placed on servers of finite capacity before workloads are realized, often with only workload means and variances known. An assignment balanced on expected load can still leave a server with more fluctuations than its residual capacity can absorb. Per-server risk constraints can enforce a prescribed overload tolerance but do not distinguish the assignments that meet it by how close the riskiest server comes to the limit. We show that, for independent tasks, placement preserves a period's total variance and total residual capacity, and decides only how fluctuation, measured by standard deviation, is paired with residual capacity across servers. The risk ratio, the ratio of fluctuation to residual capacity, parameterizes a server's worst-case overload probability over all aggregate workload distributions with given mean and variance. We minimize the largest risk ratio, attained at the risk bottleneck. Risk-Bottleneck Placement (RBP) places tasks in descending order of standard deviation, each on the server with the smallest post-assignment risk ratio, and refines the assignment with relocations and swaps within a finite per-period online budget. On small instances solved exactly by enumerating every assignment, minimizing the maximum expected utilization leaves the largest risk ratio 25.66--29.19\% above its exact minimum on average, even with risk-minimizing tie-breaking, and the corresponding gap for RBP is 1.77--2.64\%. In trace-derived multi-period simulations at two workload pressures, RBP reduces the worst-server overload probability by 26.0\% and 18.0\% relative to the best baseline and remains the lowest at every tested pressure and scale.
\end{abstract}


\section{Introduction}
\label{sec:intro}

Edge computing places computation close to users and data sources to reduce response latency, but each edge server operates with finite local capacity~\cite{ref1,ref3,ref_luo_survey}. A placement decision must often be made before the actual computation demand of the arriving tasks is realized. The processing resources 
a task requires are uncertain before execution~\cite{ref7,ref8,ref11} and may become known only when the task completes~\cite{ref11}. What the scheduler can obtain in advance are low-order statistics of each task's workload, typically its mean and variance~\cite{ref11,ref_fan_cantelli}. In an online system, the tasks of each decision period are placed using these moments together with the current server state before the workload realizations of the period are observed. A server that cannot complete its realized workload within the period carries the remainder into the next period as backlog.

\begin{figure*}[t]
    \centering
    \includegraphics[width=0.95\textwidth]{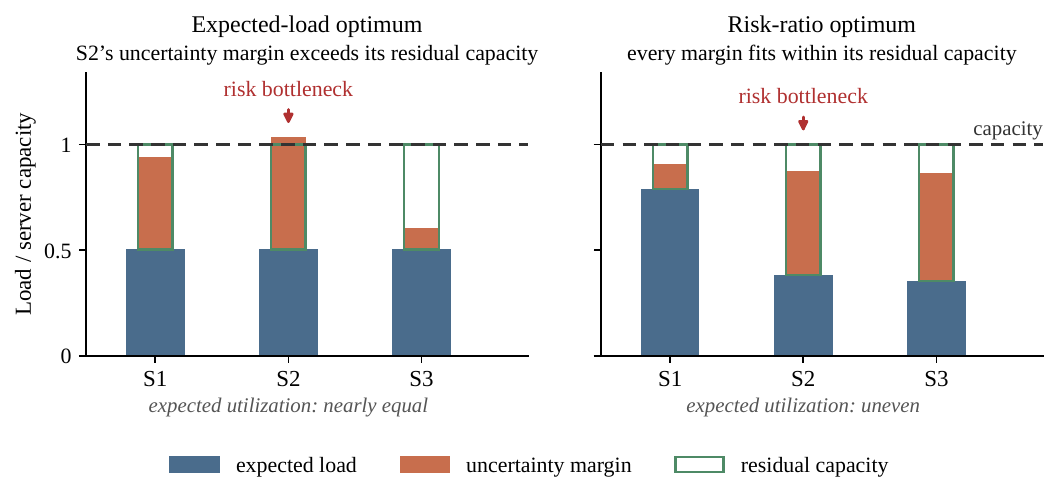}
    \caption{Expected-load optimum (left) and risk-ratio optimum (right) for one instance with 12 tasks and 3 servers, both obtained by enumerating every assignment. The expected-load optimum minimizes the maximum expected utilization and, among tied optima, minimizes the largest risk ratio. The risk-ratio optimum minimizes the largest risk ratio. Arrows mark the risk bottleneck, the server with the largest risk ratio. The uncertainty margin is a safety margin proportional to the server's workload standard deviation.}
    \label{fig:motivation}
\end{figure*}

Under this information, balancing expected workload is not the same as balancing overload risk. After a placement, each server has a residual capacity, the capacity left after its backlog and the expected workload of its new tasks, and an aggregate workload fluctuation, measured by the standard deviation of the workload it has received. The residual capacity is the headroom available to absorb that fluctuation. We call the ratio of a server's aggregate fluctuation to its residual capacity the server's \textit{risk ratio}. The larger it is, the more fluctuation the server must absorb with the headroom it has. A server can look well balanced in expectation while carrying more fluctuation than its headroom can safely absorb. Fig.~\ref{fig:motivation} illustrates this on a small instance with 12 tasks and 3 servers, for which we enumerate all $3^{12}=531{,}441$ complete assignments so that both assignments shown are exact optima. The assignment that minimizes the maximum expected utilization, which we call the expected-load optimum, gives the three servers nearly equal expected utilization. Yet S2 cannot absorb its fluctuation since its uncertainty margin, a safety margin proportional to its workload standard deviation, is larger than its residual capacity. The assignment that minimizes the largest risk ratio, the risk-ratio optimum, redistributes the same tasks so that expected utilization becomes uneven, with S1 carrying more expected load than S2 and S3, but fluctuation is better matched to the headroom of each server. The example raises a placement question that an expected-load criterion does not answer, that is, which server carries the largest fluctuation relative to the capacity left to absorb it and how a placement can make that ratio as small as possible.

Classical scheduling rules operate on deterministic task sizes or estimated execution costs~\cite{ref6}. Chance-constrained programming~\cite{ref17,ref_nemirovski} and moment-based distributionally robust optimization~\cite{ref18,ref16,ref23} provide tools for turning probabilistic requirements into deterministic constraints, and the latter can derive such constraints from moment information without assuming a particular distribution. Work on mobile edge computing with uncertain processing cycles uses these tools to bring moment information into the decision. Li et~al.~\cite{ref11} use the mean and variance of processing cycles to provide probabilistic deadline guarantees while minimizing user energy consumption, and Fan et~al.~\cite{ref_fan_cantelli} bound the computation-outage probability under an unknown workload distribution while also minimizing energy consumption. In both works, the probabilistic requirement determines which decisions are acceptable, and the optimized objective is energy consumption. Applied to placement, a requirement of this kind accepts every placement that keeps each server within the tolerance, however close its riskiest server comes to the limit. A scheduler may instead balance a risk-adjusted load, which adds an uncertainty margin to the backlog and the expected workload of each server. Keeping this load within capacity marks the same boundary as a limit on the risk ratio, but two servers with the same risk-adjusted utilization can still have different risk ratios (Section~\ref{sec:model}), so balancing risk-adjusted utilization does not directly minimize the largest one.

Under the independent-task model considered in this work, server-level overload risk has a simple structure (Section~\ref{sec:structure}). The total workload variance of a decision period is fixed by the arriving batch, and the total residual capacity is fixed by the batch and the current server state. Placement therefore cannot remove uncertainty or create headroom. It decides only how fluctuation is paired with residual capacity across servers. The risk ratio also has a distributionally robust meaning. Among all aggregate workload distributions with a server's mean and variance, the largest overload probability is an increasing function of the risk ratio, which is the tight form of Cantelli's inequality~\cite{ref23,ref18}. We call the server with the largest risk ratio the risk bottleneck and take the bottleneck risk ratio, the largest risk ratio across servers, as the objective of per-period placement. Minimizing it therefore minimizes the largest worst-case overload probability across servers. A per-server risk tolerance defines a feasibility boundary but does not distinguish the placements within it. The bottleneck risk ratio ranks them by how close their riskiest server comes to the limit. Since a placement meets the tolerance if and only if its bottleneck risk ratio does not exceed a threshold set by the tolerance (Section~\ref{sec:model}), every minimizer of the bottleneck risk ratio meets the tolerance whenever some placement does. For given residual capacities, the pairing view also shows that the bottleneck risk ratio is smallest when every server carries fluctuation in proportion to its headroom. Tasks are indivisible, so this proportion cannot always be met exactly. On small instances with 3 and 4 servers, whose exact optima we obtain by enumerating every assignment, the bottleneck risk ratio of the optimum exceeds the proportional ideal by a median of 0.33\% and 0.69\%.

Risk-Bottleneck Placement (RBP) builds on this structure in two stages that share one finite online budget.
The construction stage processes tasks in descending order of workload standard deviation and places each on the server with the smallest post-assignment risk ratio. The refinement stage then spends the remaining budget on one-task relocations and two-task swaps aimed at reducing the bottleneck risk ratio.

The main contributions are as follows.

\begin{itemize}
    \item \textbf{Pairing view of server-level overload risk.}
    Under task independence, the total workload variance and the total residual capacity of a decision period do not depend on placement, which acts only on how fluctuation is paired with headroom across servers. The \textit{risk ratio}, a server's aggregate fluctuation relative to its residual capacity, parameterizes the worst-case overload probability over all aggregate workload distributions with the given mean and variance, so minimizing the bottleneck risk ratio minimizes the largest such worst-case probability. The analysis also gives a lower bound on the bottleneck risk ratio of an assignment in terms of its residual capacities, attained if and only if every server carries fluctuation in proportion to its residual capacity.

    \item \textbf{Risk-Bottleneck Placement for finite-budget online assignment.}
    RBP assigns each task to the server with the smallest post-assignment risk ratio, with tasks taken in descending order of standard deviation. It then refines the assignment with relocations and swaps aimed at reducing the bottleneck risk ratio, within a finite online budget that includes construction.
\end{itemize}

The remainder of the paper is organized as follows. Section~\ref{sec:related} positions this work relative to scheduling under workload uncertainty. Section~\ref{sec:model} presents the system model and the risk formulation, and Section~\ref{sec:structure} derives the structural properties. Section~\ref{sec:method} describes RBP, Section~\ref{sec:eval} reports the evaluation and Section~\ref{sec:conclusion} concludes the paper.
\section{Related Work}
\label{sec:related}

\subsection{Scheduling under Uncertainty in Edge Systems}

Scheduling and load-balancing heuristics for cloud and edge systems typically place tasks according to known or predicted execution costs or resource capacities~\cite{ref6,ref28,ref29,dai2019autonomous,deng2020edgecollab}, and learning-based schedulers learn offloading and resource-allocation policies from observed system states~\cite{ref13,ref33,ref_djigal}.

A growing body of work in mobile edge computing models uncertainty explicitly, whether in offloading outcomes, in computing resources or in computation demand. Qu et~al.~\cite{ref_qu_robust} minimize the overall latency of offloaded tasks under a bounded number of uncertain offloading failures through min--max robust optimization, and Apostolopoulos et~al.~\cite{ref_apostolopoulos} study partial data offloading to UAV-mounted servers whose computing payoff is uncertain. Zhao et~al.~\cite{zhao2026cmsts} schedule task batches semi-online on MEC nodes of uncertain performance, using historical latency statistics to adjust the effective capacity of node clusters. Closer to the present setting, other studies attach the uncertainty to the computation demand of tasks. Deng et~al.~\cite{ref8} derive a closed-form expression for the average offloading success probability under uncertain processing cycles and intermittent communications and maximize this probability over offloading decisions. Statistical QoS guarantees~\cite{ref_li_qos} and moment-based probabilistic deadline and outage guarantees~\cite{ref7,ref11,ref_fan_cantelli} attach uncertainty to the requirements of individual tasks or users, whereas Chen et~al.~\cite{ref38} model uncertain task arrivals at the system level and minimize the energy-constrained expected system latency under the worst-case distribution of a data-driven ambiguity set. Execution-time uncertainty has also long been studied in real-time systems through probabilistic timing and schedulability analysis~\cite{davis2019timing,davis2019sched}, which analyzes timing distributions and deadline feasibility for a given task set and scheduling policy. Mixed-criticality scheduling associates tasks with several execution-time estimates and, when a high-criticality task overruns its lower estimate, reduces or suspends the service of lower-criticality tasks~\cite{burns2018mixed}. In our setting, the decision is the placement of a batch onto servers, and the question is how aggregate fluctuation is distributed against residual capacity across a fixed set of servers.

\subsection{Stochastic Packing, Overcommitment and Load Balancing}

Server-level capacity risk under random workloads is central to stochastic bin packing and cloud overcommitment. Cohen et~al.~\cite{cohen2019overcommitment} model overcommitment as bin packing with chance constraints, in which jobs arrive online and must be placed immediately. They show that, for independent Gaussian job sizes or for distributionally robust families with given mean and diagonal covariance, each chance constraint becomes a submodular capacity constraint, and they give online placement algorithms with constant-factor guarantees, including a competitive ratio for First-Fit, that reduce the number of machines. Their submodular formulation captures risk pooling, whereby the safety buffer required per job decreases as more jobs share a machine. In the distributionally robust case, the resulting capacity condition adds to the aggregate mean a buffer proportional to the aggregate standard deviation, the same mean-plus-buffer form as the risk-adjusted capacity condition of Section~\ref{sec:model}. Exact methods for chance-constrained bin packing include branch-and-price under stochastic and distributionally robust item sizes~\cite{zhang2020branch}. More generally, Zhang et~al.~\cite{zhang2018ambiguous} reformulate distributionally robust chance-constrained binary programs under mean--covariance information as 0--1 second-order cone programs and evaluate them on chance-constrained bin-packing instances. Yan et~al.~\cite{yan2022batch} place container batches onto machines that already carry load and minimize the cluster-wide Used Capacity at Confidence (UCaC). Under their Gaussian approximation, the UCaC of a machine is its aggregate mean plus a confidence-dependent multiple of its aggregate standard deviation, and the cluster objective sums this quantity over machines. Since the total mean over machines is fixed, minimizing the cluster UCaC amounts to minimizing the sum of per-machine standard deviations, which favors concentrating fluctuation on fewer machines. We instead minimize the largest ratio of aggregate fluctuation to residual capacity across servers, which, for given residual capacities, is smallest when fluctuation is proportional to residual capacity (Proposition~2).

In virtualized datacenters, Wu et~al.~\cite{ref_wu_hotspot} schedule virtual machines online with overcommitment, raising utilization while keeping hotspot occurrence within a target probability, and a related study uses predicted utilization and $\Gamma$-robust optimization to select live migrations that keep each physical machine hotspot-safe with a desired probability~\cite{wu2024hotspot}. In edge computing, Tang et~al.~\cite{tang2024joint} jointly learn per-node overbooking thresholds and container placement to maximize long-term provider profit, and Ma et~al.~\cite{ma2026following} maximize platform profit subject to a per-node overload-probability constraint evaluated from task usage distributions. Stochastic server consolidation assigns services or jobs with random workloads to as few servers as possible~\cite{hahnel2018extending,martinovic2022introduction}. In real-time systems, Wang et~al.~\cite{wang2017harmonicity} partition periodic tasks with probabilistic execution times onto multicore processors subject to a deadline-miss probability bound, using execution-time distributions and period harmonicity to improve schedulability. Ren et~al.~\cite{ren2019harmonic} sort probabilistic real-time tasks by workload and assign them to processors one at a time so as to minimize the increase in a harmonic index.

Stochastic load balancing shares the fixed-machine setting and a min--max objective. Kleinberg et~al.~\cite{kleinberg2000bursty} study assignments of stochastic jobs that minimize the expected maximum load and develop constant-factor approximations that summarize stochastic job sizes through effective-bandwidth measures. Stochastic load balancing has also been studied under other job-size distributions~\cite{goel1999stochastic} and on unrelated machines~\cite{gupta2021stochastic}. In parallel-machine scheduling under uncertain processing times, Song and Leus~\cite{song2022parallel} minimize the makespan under robust, chance-constrained and distributionally robust chance-constrained models, whereas Novak et~al.~\cite{novak2022scheduling} maximize the probability that all jobs with normally distributed processing times finish by a common due date.

Uncertainty-aware edge offloading optimizes quantities such as latency, energy, utility or offloading success probability, using robust, distributionally robust, probabilistic or prospect-theoretic formulations. In the stochastic packing, overcommitment and consolidation studies above, server-level capacity risk constrains placement or migration feasibility, and the objectives concern server count, utilization, UCaC, profit or migration cost, while in probabilistic real-time partitioning, execution-time distributions guide task-to-core partitioning to improve schedulability and resource utilization. Stochastic load balancing instead minimizes the expected maximum realized load, and parallel-machine scheduling under uncertainty addresses makespan under robust or chance-constrained models and the probability that all jobs meet a common due date. This paper directly optimizes a different server-level quantity. Given task moments, server capacities and the current backlog, it minimizes the largest worst-case overload probability across servers, expressed through the bottleneck risk ratio, by pairing aggregate fluctuation with residual capacity and improving that pairing within a finite online budget.
\section{System Model and Problem Formulation}
\label{sec:model}

We consider online task placement in an edge computing system where the realized computational workload of a task is unknown when the placement decision is made. A placement that balances expected load may still leave one server with more workload fluctuation than its residual capacity can absorb. We therefore model both the expected workload and the aggregate fluctuation assigned to each server, conditioned on the backlog carried from previous periods.

\subsection{Tasks and Placement}
\label{subsec:task_model}

Consider a set of heterogeneous edge servers $\mathcal{M}=\{1,\ldots,m\}$. The system operates in discrete decision periods. At the beginning of period $t$, a batch $\mathcal{B}_t$ of $n$ new tasks is available for placement.

The realized workload of task $i\in\mathcal{B}_t$ is a nonnegative random variable $\widetilde{w}_i$ with
\begin{equation}
    \mathbb{E}[\widetilde{w}_i]=\mu_i, \qquad \mathrm{Var}(\widetilde{w}_i)=\sigma_i^2 .
    \label{eq:task_moments}
\end{equation}

At decision time the scheduler knows $(\mu_i,\sigma_i)$ for every task of the current batch, but knows neither the realizations of the current batch nor any future batch. Estimation error in these moment inputs is outside the scope of this model.

Task workloads are assumed mutually independent within a decision period, which makes the variance of colocated workloads additive and is used explicitly in the structural analysis below. No parametric workload distribution is assumed.

Each task is indivisible and must be assigned to exactly one server. Let $x_{ij}(t)\in\{0,1\}$ indicate whether task $i$ is assigned to server $j$ in period $t$. A valid assignment satisfies
\begin{equation}
    \sum_{j\in\mathcal{M}} x_{ij}(t)=1, \qquad \forall i\in\mathcal{B}_t .
    \label{eq:unique_assignment}
\end{equation}

The model covers only the placement of the current batch onto a fixed set of servers with finite capacities that stay constant across periods, which corresponds to resource-constrained edge settings without elastic scaling. We do not model migration of executing work, preemption or communication delay.

\subsection{Servers and Backlog}
\label{subsec:server_model}

Server $j$ has a fixed per-period processing capacity $C_j>0$. Let $B_j(t)\ge 0$ denote the unfinished workload already present at the beginning of period $t$. This backlog is known when the current decision is made.

For an assignment $X(t)=[x_{ij}(t)]$, the realized new workload of server $j$ is
\begin{equation}
    A_j(t) = \sum_{i\in\mathcal{B}_t} x_{ij}(t)\,\widetilde{w}_i .
    \label{eq:actual_server_load}
\end{equation}

Server $j$ processes at most $C_j$ units of work during the period, and the remainder is carried forward:
\begin{equation}
    B_j(t+1) = \bigl[\,B_j(t)+A_j(t)-C_j\,\bigr]^{+},
    \label{eq:backlog_update}
\end{equation}
where $[z]^{+}=\max\{0,z\}$. Decisions in different periods are therefore coupled through backlog. Individual tasks from earlier periods are not rescheduled, and only their unfinished aggregate workload remains at the server that received them. The system is online across periods, while the tasks of one period are placed jointly as a batch.

\subsection{Aggregate Server State}
\label{subsec:aggregate_state}

For a candidate assignment $X$, the aggregate mean workload of the current batch on server $j$ is
\begin{equation}
    \mu_j(X) = \sum_{i\in\mathcal{B}_t} x_{ij}\mu_i ,
    \label{eq:server_mean}
\end{equation}
and, under the independence assumption, its aggregate variance is
\begin{equation}
    v_j(X) = \sum_{i\in\mathcal{B}_t} x_{ij}\sigma_i^2 , \qquad \sigma_j(X)=\sqrt{v_j(X)} .
    \label{eq:server_variance}
\end{equation}

The expected residual capacity, after accounting for both the existing backlog and the expected workload of the current batch, is
\begin{equation}
    d_j(X) = C_j-B_j(t)-\mu_j(X).
    \label{eq:residual_capacity}
\end{equation}
It is deterministic and equals $\mathbb{E}\bigl[C_j-B_j(t)-A_j(t)\bigr]$, and we refer to it simply as the residual capacity in the remainder of the paper.

The pair $(d_j,\sigma_j)$ is the central server state in this formulation. Here $d_j$ is the headroom available to absorb deviations above the expected workload, and the aggregate standard deviation $\sigma_j$ measures the fluctuation placed against that headroom.

\subsection{Server-Level Risk Representation}
\label{subsec:risk_representation}

Write
\begin{equation}
    p_j(X) = \Pr\bigl[\,B_j(t)+A_j(t)>C_j\,\bigr]
    \label{eq:overload_prob}
\end{equation}
for the probability that server $j$ cannot complete its assigned work within the period. When $d_j(X)>0$, Cantelli's inequality gives
\begin{equation}
    p_j(X) = \Pr\bigl[\,A_j(t)-\mu_j(X)>d_j(X)\,\bigr] \le \frac{\sigma_j^2(X)}{\sigma_j^2(X)+d_j^2(X)} .
    \label{eq:cantelli_bound}
\end{equation}

The bound depends on the aggregate fluctuation relative to the residual capacity, which motivates the server-level risk ratio
\begin{equation}
    RR_j(X) =
    \begin{cases}
        \sigma_j(X)/d_j(X), & d_j(X)>0,\\
        0, & d_j(X)=\sigma_j(X)=0,\\
        +\infty, & \text{otherwise}.
    \end{cases}
    \label{eq:risk_ratio}
\end{equation}
A server with $RR_j(X)=+\infty$ has no residual capacity left to absorb its fluctuation, whereas a server with $d_j(X)=\sigma_j(X)=0$ is loaded exactly to its capacity without fluctuation and cannot overload. For $d_j(X)>0$, the right-hand side of Eq.~\eqref{eq:cantelli_bound} equals $RR_j^2(X)/\bigl(1+RR_j^2(X)\bigr)$, which is increasing in $RR_j$. A larger $RR_j$ means that the server carries more aggregate fluctuation relative to the capacity left to absorb it.

The bound is tight when only the mean and variance of the aggregate workload are known. Let $\mathcal{P}_j(X)$ denote the set of all distributions of $A_j(t)$ with mean $\mu_j(X)$ and variance $\sigma_j^2(X)$. For $d_j(X)>0$, the worst-case overload probability over this set is
\begin{equation}
\begin{aligned}
    \overline{p}_j(X) &\triangleq \sup_{P\in\mathcal{P}_j(X)} \Pr_P\bigl[\,B_j(t)+A_j(t)>C_j\,\bigr]\\
    &= \frac{RR_j^2(X)}{1+RR_j^2(X)} ,
\end{aligned}
    \label{eq:worst_case_prob}
\end{equation}
where the supremum is approached by two-point distributions~\cite{ref23,ref18}. When $d_j(X)\le 0$, the supremum equals 0 if $d_j(X)=\sigma_j(X)=0$ and 1 otherwise, so Eq.~\eqref{eq:worst_case_prob} holds for every server with $RR_j^2/(1+RR_j^2)$ read as 1 at $RR_j=+\infty$. The risk ratio is therefore a monotone parameterization of the worst-case overload probability of server $j$ over the aggregate mean--variance ambiguity set $\mathcal{P}_j(X)$. The set $\mathcal{P}_j(X)$ constrains only the mean and variance of $A_j(t)$, whereas the task model further requires $A_j(t)$ to be a sum of independent nonnegative task workloads. The distributions that can actually arise form a subset of $\mathcal{P}_j(X)$, so $\overline{p}_j(X)$ bounds $p_j(X)$ from above but need not be attained by the task workloads themselves.

The bottleneck risk ratio of an assignment is
\begin{equation}
    RR_{\max}(X) = \max_{j\in\mathcal{M}} RR_j(X),
    \label{eq:rr_max}
\end{equation}
and a server attaining it is a risk bottleneck. Since $\overline{p}_j$ is increasing in $RR_j$, the risk bottleneck is also a server with the largest worst-case overload probability and minimizing $RR_{\max}(X)$ minimizes $\max_{j}\overline{p}_j(X)$.

Expected-load balancing asks how full a server is on average, as measured by the expected utilization
\begin{equation}
    u_j(X)=\frac{B_j(t)+\mu_j(X)}{C_j}.
    \label{eq:exp_util}
\end{equation}
A risk bottleneck, by contrast, is a server whose aggregate fluctuation is largest relative to its residual capacity.

Let $\alpha\in(0,1)$ be the prescribed per-server risk tolerance and
\begin{equation}
    \kappa = \sqrt{\frac{1}{\alpha}-1}\,.
    \label{eq:kappa}
\end{equation}

The risk-adjusted load of server $j$ is
\begin{equation}
    \widehat{L}_j(X) = B_j(t)+\mu_j(X)+\kappa\sigma_j(X),
    \label{eq:risk_adjusted_load}
\end{equation}
where $\kappa\sigma_j(X)$ is the uncertainty margin added to the backlog and the expected workload.

For $d_j(X)>0$, the risk-adjusted utilization and the risk ratio are related by
\begin{equation}
    \frac{\widehat{L}_j(X)}{C_j} = 1-\frac{d_j(X)}{C_j}\bigl(1-\kappa\,RR_j(X)\bigr),
    \label{eq:util_vs_rr}
\end{equation}
so the risk-adjusted capacity condition $\widehat{L}_j(X)\le C_j$ holds if and only if $RR_j(X)\le 1/\kappa$. When $d_j(X)\le 0$, each of the two conditions holds if and only if $d_j(X)=\sigma_j(X)=0$, so the equivalence holds for every server. Substituting $RR_j(X)\le 1/\kappa$ into Eq.~\eqref{eq:worst_case_prob} gives
\begin{equation}
    p_j(X) \le \overline{p}_j(X) \le \frac{1/\kappa^2}{1+1/\kappa^2} = \frac{1}{1+\kappa^2} = \alpha .
    \label{eq:overload_guarantee}
\end{equation}

The equivalence concerns feasibility, not the ordering of feasible servers. Inside the boundary, the risk-adjusted utilization also depends on the normalized residual capacity $d_j(X)/C_j$, so two servers with the same risk-adjusted utilization can have different risk ratios, and an assignment that balances risk-adjusted utilization need not minimize $RR_{\max}(X)$.

An assignment is \emph{risk-feasible} when every server satisfies the risk-adjusted capacity condition, equivalently when $RR_{\max}(X)\le 1/\kappa$. The worst-case overload probability $\overline{p}_j(X)$ is then at most $\alpha$ on every server.

\subsection{Problem Formulation}
\label{subsec:problem_formulation}

Let
\begin{equation}
    \mathcal{F}_t = \bigl\{\,X:\ \widehat{L}_j(X)\le C_j,\ \forall j\in\mathcal{M} \,\bigr\}
    \label{eq:feasible_set}
\end{equation}
denote the set of risk-feasible assignments of the current batch. The placement problem in period $t$ is
\begin{equation}
\begin{aligned}
    \min_{X} \quad & RR_{\max}(X)\\
    \text{s.t.}\quad & \sum_{j\in\mathcal{M}}x_{ij}=1, && \forall i\in\mathcal{B}_t,\\
    & x_{ij}\in\{0,1\}, && \forall i\in\mathcal{B}_t,\; \forall j\in\mathcal{M}.
\end{aligned}
\label{eq:problem}
\end{equation}

The problem asks how the current batch should be placed so that the risk bottleneck carries as little fluctuation as possible relative to its residual capacity. By Eq.~\eqref{eq:worst_case_prob}, this minimizes the largest worst-case overload probability across servers. The risk tolerance does not need to be imposed as a separate constraint. Since $X\in\mathcal{F}_t$ if and only if $RR_{\max}(X)\le 1/\kappa$ (Section~\ref{subsec:risk_representation}), every optimal solution of Eq.~\eqref{eq:problem} is risk-feasible whenever $\mathcal{F}_t$ is nonempty. The tolerance $\alpha$ therefore determines whether an assignment can be certified at level $\alpha$, but not which assignment minimizes $RR_{\max}$.

Backlog and batch demand can leave $\mathcal{F}_t$ empty, in which case no assignment meets the tolerance. Since the model admits neither task rejection nor deferral, RBP still returns a complete assignment in such periods (Section~\ref{sec:method}), and Section~\ref{sec:eval} reports how often the returned assignments are risk-feasible.

Problem~\eqref{eq:problem} is NP-hard. With two servers of capacity $C$, no backlog and tasks whose mean and variance both equal $a_i>0$, where $\Lambda=\sum_{i\in\mathcal{B}_t}a_i<C$, the risk ratio $RR_j(X)=\sqrt{\mu_j(X)}/\bigl(C-\mu_j(X)\bigr)$ increases with $\mu_j(X)$, so $RR_{\max}(X)\le\sqrt{\Lambda/2}/(C-\Lambda/2)$ holds if and only if $X$ splits the $a_i$ into two sets of equal sum, which is the NP-complete PARTITION problem.

\section{Structural Properties}
\label{sec:structure}

The risk representation above has a structure that expected load alone does not show. Let
\begin{equation}
    V_t = \sum_{i\in\mathcal{B}_t}\sigma_i^2, \qquad
    D_t = \sum_{j\in\mathcal{M}} C_j - \sum_{j\in\mathcal{M}} B_j(t) - \sum_{i\in\mathcal{B}_t}\mu_i .
    \label{eq:totals}
\end{equation}

\subsection{Conservation under Placement}

\noindent\textbf{Proposition 1.}
Under the independent-task model, every complete assignment $X$ satisfies
\begin{equation}
    \sum_{j\in\mathcal{M}}v_j(X)=V_t, \qquad \sum_{j\in\mathcal{M}}d_j(X)=D_t .
    \label{eq:conservation}
\end{equation}

\noindent\textit{Proof.}
Since every task is assigned to exactly one server, $\sum_j v_j(X)=\sum_j\sum_i x_{ij}\sigma_i^2=\sum_i\sigma_i^2=V_t$, and likewise $\sum_j d_j(X)=\sum_j\bigl(C_j-B_j(t)\bigr)-\sum_j\mu_j(X)=\sum_j C_j-\sum_j B_j(t)-\sum_i\mu_i=D_t$. \hfill$\square$

By Proposition~1, placement can neither reduce the total variance the model attributes to the batch nor create residual capacity. It determines only how aggregate fluctuation is paired with residual capacity across servers, and that pairing is the structural source of server-level risk imbalance.

\subsection{Proportional Risk Balancing}

The same structure yields an ideal lower bound on the bottleneck risk ratio for a given residual-capacity vector.

\noindent\textbf{Proposition 2.}
For an assignment $X$ with $d_j(X)>0$ on every server,
\begin{equation}
    RR_{\max}(X) \;\ge\; I(X) \;\triangleq\; \sqrt{\frac{V_t}{\sum_{j\in\mathcal{M}}d_j^2(X)}}\,,
    \label{eq:conditional_ideal}
\end{equation}
with equality if and only if
\begin{equation}
    \frac{\sigma_1(X)}{d_1(X)} = \cdots = \frac{\sigma_m(X)}{d_m(X)} .
    \label{eq:proportional_condition}
\end{equation}

\noindent\textit{Proof.}
Let $r=\max_j\sigma_j(X)/d_j(X)=RR_{\max}(X)$, so that $\sigma_j(X)\le r\,d_j(X)$ for every $j$. By Proposition~1, $V_t=\sum_j\sigma_j^2(X)\le r^2\sum_j d_j^2(X)$, which is Eq.~\eqref{eq:conditional_ideal}. Equality requires $\sigma_j^2(X)=r^2d_j^2(X)$ for every $j$. As $\sigma_j(X)$, $r$ and $d_j(X)$ are nonnegative, this is $\sigma_j(X)=r\,d_j(X)$, that is, Eq.~\eqref{eq:proportional_condition}. \hfill$\square$

The quantity $I(X)$ depends on the residual capacities produced by the assignment itself, so it lower-bounds the bottleneck risk ratio of that assignment rather than the optimum of the discrete problem. By Eq.~\eqref{eq:proportional_condition}, the bottleneck risk ratio reaches this bound if and only if the fluctuation assigned to each server is proportional to the residual capacity available there. We refer to this proportionality as risk balance. Since tasks are indivisible and their means and variances move together, a scheduler cannot in general enforce that proportion exactly.

\subsection{Implication for Bottleneck-Oriented Search}

The bottleneck objective also restricts which servers can produce an immediate improvement. Suppose server $b$ is the unique risk bottleneck under $X$, that is, $RR_b(X)>RR_j(X)$ for all $j\neq b$. A relocation of one task or a swap of two tasks that leaves server $b$ untouched changes the risk ratios of at most two other servers and leaves $RR_b$ as it was, so the maximum cannot fall. Any one-step strict reduction of a unique bottleneck risk ratio must therefore involve the current bottleneck. This one-step property motivates prioritizing candidates involving the current risk bottleneck under a limited online search budget.
\section{Risk-Bottleneck Placement}
\label{sec:method}

The model and structural analysis in Sections~\ref{sec:model} and~\ref{sec:structure} show that server risk depends not only on expected load but also on how aggregate workload fluctuation is matched to the residual capacity of each server. Two properties guide the design of RBP. First, when the maximum bottleneck risk ratio is attained by a unique server, a single move that strictly reduces it must modify that server. Second, a placement, relocation or swap changes the aggregate state of at most two servers while all other server states remain unchanged.

Risk-Bottleneck Placement (RBP) uses these properties in two stages. It first constructs a complete assignment by processing tasks with larger workload standard deviation earlier and placing each task on the server with the smallest post-assignment risk ratio. It then spends the remaining online budget refining that assignment with relocations and swaps, evaluated first on the servers with the largest risk ratios. Candidate moves are evaluated incrementally from the servers they modify. All decisions are based on standard deviations and risk ratios, so the risk tolerance $\alpha$ does not enter the algorithm. It only determines whether the returned assignment is risk-feasible. Algorithm~\ref{alg:rbp} summarizes the procedure.

\begin{algorithm}[tb]
\caption{Risk-Bottleneck Placement}
\label{alg:rbp}
\begin{algorithmic}[1]
\REQUIRE Task batch $\mathcal{B}_t$; server states $\mathcal{S}_t=\{B_j(t),C_j\}_{j\in\mathcal{M}}$; online budget $T_{\max}$
\ENSURE Assignment $X$
\STATE $\tau\gets\textsc{Now}()+T_{\max}$ \COMMENT{deadline}
\STATE $X\gets\textsc{RiskGuidedConstruction}(\mathcal{B}_t,\mathcal{S}_t)$
\WHILE{$\textsc{Now}()<\tau$}
    \STATE $\mathrm{committed}\gets\textsc{BottleneckRefine}(X,\mathcal{B}_t,\mathcal{S}_t,\tau)$
    \IF{\textbf{not} $\mathrm{committed}$}
        \STATE \textbf{break}
    \ENDIF
\ENDWHILE
\RETURN $X$
\end{algorithmic}
\end{algorithm}

\subsection{Risk-Guided Construction}
\label{subsec:construction}

RBP sorts the tasks in non-increasing order of their standard deviations $\sigma_i$ and places them one at a time. For a candidate server $j$, let $\mu_j$ and $v_j$ denote the aggregate mean and variance of the tasks already assigned to it in the current batch. If task $i$ is tentatively placed on $j$,
\begin{equation}
\begin{aligned}
\mu'_j&=\mu_j+\mu_i, & v'_j&=v_j+\sigma_i^2,\\
\sigma'_j&=\sqrt{v'_j}, & d'_j&=C_j-B_j(t)-\mu'_j ,
\end{aligned}
\end{equation}
which give the post-assignment risk ratio $RR'_j$ of the receiving server through Eq.~\eqref{eq:risk_ratio}. RBP places the task on the server with the smallest $RR'_j$. Algorithm~\ref{alg:construction} gives the construction procedure.

\begin{algorithm}[tb]
\caption{Risk-Guided Construction}
\label{alg:construction}
\begin{algorithmic}[1]
\REQUIRE Task batch $\mathcal{B}_t$; server states $\mathcal{S}_t$
\ENSURE Assignment $X$
\STATE $\mu_j\gets0,\;v_j\gets0,\quad\forall j\in\mathcal{M}$
\STATE Sort tasks by non-increasing $\sigma_i$
\FOR{each task $i$ in the sorted order}
    \FOR{each server $j\in\mathcal{M}$}
        \STATE Compute $\mu'_j$, $v'_j$, $\sigma'_j$, $d'_j$ and $RR'_j$
    \ENDFOR
    \STATE $j^\ast\gets\arg\min_{j\in\mathcal{M}}RR'_j$
    \STATE Assign task $i$ to $j^\ast$; update $\mu_{j^\ast}$ and $v_{j^\ast}$
\ENDFOR
\RETURN $X$
\end{algorithmic}
\end{algorithm}

\noindent\textbf{Ordering.}
A task with a larger $\sigma_i$ contributes more variance to whichever server receives it, so its placement can have a larger effect on the resulting risk profile. Processing such tasks earlier gives these higher-impact decisions access to a larger set of remaining server states.

\medskip
\noindent\textbf{Risk feasibility of the receiver.}
A single insertion changes only the aggregate state of the receiving server, and by Section~\ref{subsec:risk_representation} the receiver satisfies the risk-adjusted capacity condition after the insertion if and only if $RR'_j\le 1/\kappa$. Whenever some server can receive the task within this condition, the server with the smallest $RR'_j$ is therefore one of them, and the receiver rule needs no separate feasibility check. When no server can, the task is still placed on the server with the smallest $RR'_j$. The assignment then lies outside $\mathcal{F}_t$, and refinement continues to reduce its bottleneck risk ratio.

\medskip
\noindent\textbf{Alignment with the bottleneck objective.}
For nonnegative task demand, inserting a task cannot decrease the risk ratio of its receiver. Let $r_{\mathrm{cur}}$ denote the largest server risk ratio before the insertion. If server $j$ receives the task, the resulting system maximum is $\max\{r_{\mathrm{cur}},\,RR'_j\}$. Since $r_{\mathrm{cur}}$ is common to all candidate receivers, minimizing $RR'_j$ also minimizes the bottleneck risk ratio attained immediately after the insertion. Comparing $RR'_j$ directly additionally distinguishes candidate receivers when several of them lie below the current maximum.

\medskip
\noindent\textbf{Relation to a risk-adjusted-utilization rule.}
A receiver rule that minimizes the post-assignment risk-adjusted utilization $\widehat{L}'_j/C_j$ also selects a receiver within the risk-adjusted capacity condition whenever one exists, since $\widehat{L}'_j/C_j\le1$ if and only if that condition holds. The two rules share this boundary, but by Eq.~\eqref{eq:util_vs_rr} they can rank the receivers inside it differently, because risk-adjusted utilization also depends on the normalized residual capacity $d'_j/C_j$. The SG-$\sigma$ baseline of Section~\ref{sec:eval} uses the same task ordering as RBP with the risk-adjusted-utilization rule, so the two constructions differ only in how they rank receivers.

\subsection{Bottleneck-Focused Refinement}
\label{subsec:refinement}

Sequential construction produces a complete assignment, but later placements can change which server is the risk bottleneck. RBP therefore spends the remaining online budget refining the constructed assignment. Instead of scanning local moves in an arbitrary order, it evaluates first the servers and tasks with the largest potential effect on the current bottleneck risk ratio.

\medskip
\noindent\textbf{Acceptance criterion.}
RBP compares candidate assignments lexicographically by
\begin{equation}
\Phi(X)=\Bigl( \max_{j\in\mathcal{M}}\overline{p}_j(X),\; \sum_{j\in\mathcal{M}}\overline{p}_j(X) \Bigr),
\label{eq:acceptance}
\end{equation}
where $\overline{p}_j(X)$ is the worst-case overload probability of Eq.~\eqref{eq:worst_case_prob}. Since $\overline{p}_j$ is increasing in $RR_j$ and $X\in\mathcal{F}_t$ if and only if $RR_{\max}(X)\le1/\kappa$, the first component already ranks every risk-feasible assignment ahead of every risk-infeasible one, so no separate feasibility term is needed. A candidate is a \emph{strict} improvement when it reduces $RR_{\max}$. Otherwise, it can be preferred only by reducing the second term at an unchanged bottleneck risk ratio. That term only distinguishes plateau candidates.

\begin{algorithm}[tb]
\caption{\textsc{BottleneckRefine}: One Refinement Step}
\label{alg:refine}
\begin{algorithmic}[1]
\REQUIRE Current assignment $X$ and its server aggregates; task batch $\mathcal{B}_t$; server states $\mathcal{S}_t$; deadline $\tau$
\ENSURE \texttt{true} if one move is committed, \texttt{false} otherwise
\FOR{each source server $a$ in \textsc{SourceOrder}$(X)$}
    \IF{$a$ contains no task from $\mathcal{B}_t$}
        \STATE \textbf{continue}
    \ENDIF
    \STATE $\mathrm{plateau}\gets\varnothing$
    \STATE Order the tasks on $a$ by non-increasing removal gain $g_i$
    \STATE \textit{Relocation stage}
    \FOR{each task $i$ in that order}
        \FOR{each destination $b\neq a$}
            \STATE Evaluate moving $i$ from $a$ to $b$ \COMMENT{keep it as $\mathrm{plateau}$ if it improves $\Phi$ but is not strict}
            \STATE Periodically, if $\textsc{Now}()\ge\tau$, \textbf{return} \texttt{false}
        \ENDFOR
        \IF{a strict candidate was found for $i$}
            \STATE Commit the best such candidate; \textbf{return} \texttt{true}
        \ENDIF
    \ENDFOR
    \STATE \textit{Swap stage}
    \FOR{each task $i$ in the same order}
        \FOR{each destination $b\neq a$ containing at least one task}
            \STATE $h\gets$ the minimum-$\sigma$ task on $b$
            \STATE Evaluate exchanging $i$ and $h$ \COMMENT{keep it as $\mathrm{plateau}$ if it improves $\Phi$ but is not strict}
            \STATE Periodically, if $\textsc{Now}()\ge\tau$, \textbf{return} \texttt{false}
        \ENDFOR
        \IF{a strict candidate was found for $i$}
            \STATE Commit the best such candidate; \textbf{return} \texttt{true}
        \ENDIF
    \ENDFOR
    \IF{$\mathrm{plateau}\neq\varnothing$}
        \STATE Commit $\mathrm{plateau}$; \textbf{return} \texttt{true}
    \ENDIF
\ENDFOR
\RETURN \texttt{false}
\end{algorithmic}
\end{algorithm}

Each invocation of Algorithm~\ref{alg:refine} commits at most one move. After a move is committed, the next invocation recomputes the server order from the updated assignment.

\medskip
\noindent\textbf{Source-server selection.}
Source servers are visited in decreasing order of $RR_j$. Since a server violates the risk-adjusted capacity condition if and only if $RR_j>1/\kappa$, servers that are not risk-feasible are visited first, the most severe violations earliest, followed by the remaining servers with the largest risk ratios.

\medskip
\noindent\textbf{Victim ordering.}
For each task $i$ assigned to source server $a$, RBP computes the decrease in the worst-case overload probability of the source obtained by removing that task,
\begin{equation}
g_i = \overline{p}_a(X)-\overline{p}_a\bigl(X\setminus\{i\}\bigr),
\label{eq:removal_gain}
\end{equation}
and examines the tasks in non-increasing order of $g_i$. Under a finite budget, this ordering determines which candidates are reached first and thus where the search effort goes.

\medskip
\noindent\textbf{Relocation.}
For task $i$ on source server $a$, every server $b\neq a$ is examined as a destination. Within one victim, RBP evaluates all destinations and retains the best strict candidate according to~\eqref{eq:acceptance}. The move is committed for the first victim in removal-gain order that admits a strict relocation. The search is therefore first-improvement across victims and best-improvement among destinations for a fixed victim.

\medskip
\noindent\textbf{Swap.}
A relocation transfers the complete mean and variance contribution of task $i$ to the destination and may therefore fail when no receiver can absorb the task without creating a worse state. A swap instead exchanges $i$ with a resident task $h$ of the destination, so the destination's aggregate variance changes by $\sigma_i^2-\sigma_h^2$ rather than by $\sigma_i^2$. RBP considers swaps only after no strict relocation is available for the current source.

For each candidate destination, only its resident task with the smallest standard deviation is considered as the swap partner. This limits the pairwise neighborhood to one swap candidate per destination and avoids returning a highly variable task to the source server that the search is attempting to relieve.

\medskip
\noindent\textbf{Plateau handling.}
While relocation and swap candidates are evaluated, RBP also records candidates that are not strict but improve the comparison in~\eqref{eq:acceptance}. Such a candidate lowers the sum of per-server worst-case overload probabilities while leaving the bottleneck risk ratio unchanged. The best plateau candidate is committed only when the current source provides neither a strict relocation nor a strict swap. Secondary improvement therefore does not take priority over an available strict reduction of the bottleneck risk ratio.

\subsection{Incremental Evaluation}
\label{subsec:incremental}

A relocation or swap changes the aggregate mean and variance of two servers and leaves all other servers unchanged. RBP therefore maintains, for every server, its aggregate mean, variance, standard deviation, residual capacity and risk ratio. A candidate is evaluated by recomputing these quantities for the two affected servers while reusing the maintained state of the remaining servers. After a move is committed, the affected server states and task-membership information are updated before the next refinement step. Ties among candidate servers and tasks are broken by their indices.

\subsection{Online Execution}
\label{subsec:budget}

Construction and refinement share a single finite online budget $T_{\max}$ per decision period, which limits the solver-side computation for one placement decision. We set $T_{\max}=25$~ms by default to test whether a placement decision can be computed within a few tens of milliseconds, and Section~\ref{subsec:runtime} shows that the worst-server overload probability of RBP varies by at most 1.27\% across budgets from 5 to 100~ms.

Deadline checking is cooperative rather than preemptive. Construction runs to completion, and refinement checks the deadline periodically during candidate scanning. When the deadline is reached, RBP discards the candidate under evaluation and returns the current complete assignment.

\subsection{Complexity}
\label{subsec:complexity}

Sorting the batch by $\sigma_i$ costs $O(n\log n)$, and evaluating all $m$ candidate receivers for each of the $n$ tasks costs $O(nm)$.

During refinement, the servers are ordered by their current risk ratios, and the tasks of each visited source server are ordered by their removal gains. Across all visited sources, each task can be examined as a victim, and each victim considers at most $m-1$ relocation destinations and, when no strict relocation is available for the current source, at most $m-1$ swap destinations. A complete refinement scan therefore examines at most $O(nm)$ relocation and swap candidates, in addition to the ordering costs. Additional state updates are required only after a move is committed. The actual number of refinement steps executed in each decision period is limited by the finite online time budget.
\section{Performance Evaluation}
\label{sec:eval}

We evaluate RBP from four perspectives. On small instances whose exact optima are obtained by enumerating every assignment, we test whether the structure identified in Section~\ref{sec:structure} appears in exact risk-ratio optima and measure how closely each online method approaches the exact optimum. The main multi-period experiments compare RBP with expected-load and risk-aware baselines at two workload pressures. Ablations quantify what refinement adds to the risk-guided construction and what the bottleneck-focused visiting order adds within refinement. Load, scale and budget experiments examine how the observed behavior changes outside the default condition.

Each experiment family draws its own instances and random streams. Within a family all methods see identical instances, identical capacities and identical evaluation streams, so the comparisons reported below are made within a family, and absolute values are not compared across families.

\subsection{Experimental Setup}
\label{subsec:setup}

\subsubsection{Trace-Derived Workload Construction}

We use Google Cluster Trace 2019~\cite{google_trace,ref_tirmazi} to construct empirical task-level workload profiles rather than to replay a datacenter workload timeline. The trace supplies the heterogeneity of workloads across tasks and the empirical distribution of each task's workload, while the batch process, server capacities, backlog evolution and placement decisions follow the model of Section~\ref{sec:model} and the configuration below. We take the job-instance usage records of cell~a over hours 24 to 48 of the trace. Each task is one job instance, and each observation is its average normalized CPU usage over a complete 300-second window. Since all windows have the same duration, an observation is proportional to the work the task performs in its window, and one decision period therefore represents a five-minute workload interval. We retain the tasks with at least 12 such observations and a positive mean and variance, and the retained observations of a task define its empirical workload distribution. Workloads, capacities and backlog are expressed in a common normalized work unit.

Some tasks in the retained pool cannot satisfy the risk-adjusted capacity condition on any server at the capacities calibrated by Eq.~\eqref{eq:capacity_calibration}, even when placed alone. A batch that contains such a task admits no risk-feasible assignment. To exclude such tasks, we remove the tasks with the largest risk-adjusted size $\mu_i+\kappa\sigma_i$. The removed fraction was fixed before the formal evaluation by a rule that involves no placement method. Among prespecified fractions, it is the smallest for which, in each of the eight combinations of $m\in\{5,10,20,50\}$ and $\rho_R\in\{0.90,0.95\}$, at most 1\% of 1{,}000 sampled batches contain such a task. The rule selects 2\%. The removed tasks account for 28.6\% of the total mean and 78.6\% of the total variance of the unfiltered pool, and 954{,}211 tasks remain. Capacities are calibrated by Eq.~\eqref{eq:capacity_calibration} on batches drawn from the retained pool, so the workload pressure after trimming is still set by $\rho_R$.

Unless stated otherwise, each period samples a batch of $n=100$ tasks from this pool without replacement, and all methods receive the same batch. The exact-small experiments use smaller batches from the same pool. All placement methods observe only the moments $(\mu_i,\sigma_i)$ at decision time, as assumed in Section~\ref{sec:model}, and never the realized workload of the current period. The realized workload of a task is drawn uniformly with replacement from its retained observations, independently across tasks, and both the Monte Carlo samples used to estimate overload probabilities and the realized path used to update backlog are generated in this way. Since $\mu_i$ and $\sigma_i^2$ are the mean and variance of the same observations, the scheduler receives the exact moments of this resampling distribution, and the experiments introduce no moment-estimation error. This construction preserves the empirical workload distribution of each retained task while matching the independent-task model of Sections~\ref{sec:model} and~\ref{sec:structure}. Cross-task correlations in the original trace are not reproduced.

\subsubsection{System Configuration}

The default system contains $m=10$ heterogeneous servers and runs for $T=100$ consecutive periods, with backlog propagated by Eq.~\eqref{eq:backlog_update} from an empty initial state. Server capacities are generated once at the beginning of a run and stay fixed for that run. Relative capacities are perturbed independently within $\pm10\%$ and normalized to preserve the prescribed total.

Workload pressure is controlled through the robust load ratio $\rho_R$. For the $T$ batches of one run, the total capacity is calibrated as
\begin{equation}
    \sum_{j\in\mathcal{M}}C_j
    =
    \frac{\displaystyle\frac{1}{T}\sum_{t=1}^{T}\left[\sum_{i\in\mathcal{B}_t}\mu_i+\kappa\sqrt{m\sum_{i\in\mathcal{B}_t}\sigma_i^2}\right]}{\rho_R}.
    \label{eq:capacity_calibration}
\end{equation}
The bracketed term is the risk-adjusted demand of a batch when its total variance $V_t$ is split evenly across the $m$ servers, since each server then carries an uncertainty margin $\kappa\sqrt{V_t/m}$ and the margins sum to $\kappa\sqrt{mV_t}$. The robust load ratio $\rho_R$ is thus the ratio of this demand, averaged over the run, to the total capacity, and a larger $\rho_R$ leaves less spare capacity for the same workload, that is, higher workload pressure. The calibration is performed before the run and gives all methods the same capacities. Within the run, each placement decision uses only the current batch and the current server state. The exact-small instances consist of a single batch and are calibrated by the same expression with $T=1$ and $\rho_R=0.90$.

Table~\ref{tab:params} summarizes the default configuration. The two main conditions use $\rho_R\in\{0.90,0.95\}$. The risk tolerance is fixed at $\alpha=0.15$ before the evaluation runs, which gives $\kappa=2.3805$ through Eq.~\eqref{eq:kappa}. The same $\kappa$ is used by the three risk-aware baselines described below, RG, SG-$\sigma$ and RA-FF, and by the capacity calibration of Eq.~\eqref{eq:capacity_calibration}, whereas RBP does not use $\alpha$ in its placement decisions (Section~\ref{sec:method}). Unless stated otherwise, each condition contains 30 independent runs, and RBP uses the 25~ms finite online budget of Section~\ref{subsec:budget}.

\begin{table}[t]
\centering
\caption{Default experimental configuration.}
\label{tab:params}
\begin{tabular}{lll}
\toprule
Parameter & Symbol & Value \\
\midrule
Number of servers & $m$ & 10 \\
Capacity heterogeneity & $C_j$ & $\pm10\%$ around the mean \\
Batch size per period & $n$ & 100 \\
Period length & -- & 300~s \\
Periods per run & $T$ & 100 \\
Independent runs & $R$ & 30 \\
Target robust load ratio & $\rho_R$ & 0.90, 0.95 \\
Risk tolerance & $\alpha$ & 0.15 \\
Robustness coefficient & $\kappa$ & 2.3805 \\
Monte Carlo samples per period & $S$ & 10{,}000 \\
Online budget of RBP & $T_{\max}$ & 25~ms \\
\bottomrule
\end{tabular}
\end{table}

\subsubsection{Baselines}

We compare RBP with four rule-based baselines and one auxiliary local-search method. Since closely related published methods differ in objective or decision model (Section~\ref{sec:related}), the baselines are chosen to separate the placement principles relevant to the present problem. DG represents expected-load balancing, as in load-balancing heuristics that place tasks according to known or predicted execution costs or resource capacities~\cite{ref28,ref29}. RG and SG-$\sigma$ use the mean-plus-standard-deviation risk adjustment that underlies moment-based capacity conditions~\cite{cohen2019overcommitment,ref11}, and SG-$\sigma$ additionally shares the task order of RBP. RA-FF adapts the ordering and first-fit structure of~\cite{ref11}, and Exp-LS is an auxiliary control that applies additional search to an expected-load objective. All of them observe the same task information and server state, namely the moments of the current batch, the current backlog and the server capacities, and none observes a realization. Each rule below is described by the order in which it processes tasks, the criterion by which it selects a receiver, the feasibility condition it applies and its behavior when no receiver satisfies that condition.

\begin{itemize}
\item \textbf{DG (deterministic greedy).} Tasks are processed in input order, and each task is placed on the server with the smallest post-assignment expected utilization $(B_j+\mu_j+\mu_i)/C_j$. DG does not read $\sigma_i$ and applies no feasibility condition.

\item \textbf{RG (robust greedy).} Tasks are processed in input order, and each task is placed on the server with the smallest post-assignment risk-adjusted utilization $\widehat{L}'_j/C_j$. No feasibility condition is applied.

\item \textbf{SG-$\sigma$.} Tasks are processed in non-increasing order of $\sigma_i$, and the receiver criterion is that of RG. SG-$\sigma$ shares its task ordering with the construction of RBP and differs from it only in the receiver rule, so it shows what the $\sigma$ ordering provides without the receiver rule and the refinement stage of RBP.

\item \textbf{RA-FF (risk-adjusted first-fit).} This baseline adapts the ordering and first-fit placement of the energy-minimized solution with probabilistic deadline guarantee (EPD) in~\cite{ref11}. Tasks are ordered by non-decreasing $\mu_i+\kappa\sigma_i$, which is the ordering key of~\cite{ref11} without its task-deadline normalization, since this model has no task-level deadlines. Each task is placed on the first server, scanned by index, that satisfies the risk-adjusted capacity condition $\widehat{L}'_j\le C_j$ after the insertion, as in the first-to-fit placement of~\cite{ref11}. If no server satisfies the condition, the task goes to the one with the smallest resulting risk-adjusted load, since tasks cannot be rejected. The risk-adjusted capacity condition thus serves only as a feasibility test and does not rank the servers that pass it.

\item \textbf{Exp-LS.} This method minimizes the maximum expected utilization $\max_j u_j(X)$ by local search, starting from an expected-load construction, within a soft budget of 100~ms, compared with the 25~ms default of RBP. We include it to examine whether search effort under the expected-load objective produces the same risk behavior.
\end{itemize}

\subsubsection{Metrics and Evaluation Protocol}

The primary metric is the empirical counterpart of the bottleneck objective of Eq.~\eqref{eq:rr_max}. For each period we draw $S=10{,}000$ workload realizations by independently resampling each task's trace-derived workload profile and estimate the overload probability of server $j$ as
\begin{equation}
    \widehat{p}_j(t)=\frac{1}{S}\sum_{s=1}^{S}\mathbf{1}\!\left[B_j(t)+\sum_{i\in\mathcal{B}_t}x_{ij}(t)\,\widetilde{w}_i^{(s)}>C_j\right].
    \label{eq:empirical_overload}
\end{equation}
The period-level value is
\begin{equation}
    \widehat{p}_{\max}(t)=\max_{j\in\mathcal{M}}\widehat{p}_j(t),
    \label{eq:empirical_worst_overload}
\end{equation}
and we average it over the 100 periods of a run. The maximum is taken across servers within each period before averaging over periods, so the metric evaluates the worst-server overload outcome corresponding to the bottleneck objective. The objective minimizes the largest worst-case overload probability over the aggregate mean--variance ambiguity sets of Section~\ref{subsec:risk_representation}, and the primary metric tests whether this also lowers the worst-server overload probability under the trace-derived workload distributions.

We also report the probability that at least one server overloads in a period, the fraction of periods whose returned assignment belongs to $\mathcal{F}_t$, the normalized system backlog $\sum_j B_j(t)/\sum_j C_j$ measured before the decision and the largest normalized single-server backlog. The risk-feasible rate records how often the moment-based guarantee of Eq.~\eqref{eq:overload_guarantee} certifies the returned assignment at level $\alpha$ for every aggregate distribution with the given moments. The two overload probabilities are estimated from the Monte Carlo samples, whereas backlog evolves along one realized path through Eq.~\eqref{eq:backlog_update}. Overload probabilities and backlog thus measure different consequences of the same decision.

Runtime is wall-clock time over the complete solver call. Since construction runs to completion and refinement checks the deadline only periodically (Section~\ref{subsec:budget}), we report the median, the 99th percentile, the maximum and the observed fraction of periods exceeding 25~ms.

All methods within the same run and period use the same sampled task batch, workload realizations and evaluation streams. Per-period values are averaged within a run, and reported entries are means over the 30 runs. The ``$\pm$'' term shown for the primary metric is the standard deviation across runs. For the primary metric in the main experiment, we compare RBP with RG, SG-$\sigma$ and RA-FF on the 30 paired run-level values at each of the two pressures. Each comparison uses a two-sided Wilcoxon signed-rank test, and the six $p$-values are adjusted with the Holm procedure. A difference is called statistically significant when its adjusted $p$-value is below 0.05. The protocol, the solver, the baselines, the metrics, the statistical tests and the seed derivation were frozen before these runs, and the seeds are disjoint from those used during development. All methods are implemented in Python 3.10 with NumPy and executed in a single process on the same machine, which has an AMD Ryzen 9 8945HX processor and 32 GB of RAM and runs Windows 11.

\subsection{Structure and Optimality on Enumerated Small Instances}
\label{subsec:exact_small}

Small instances allow us to answer two questions that the main scale cannot. The first is whether the structure derived in Section~\ref{sec:structure} appears in exact risk-ratio optima, and the second is how far each online method remains from that optimum. Each instance has $n=12$ tasks and $m\in\{3,4\}$ servers with no initial backlog, with 20 held-out instances per server count. Every complete assignment is enumerated, which yields the exact minima of both $RR_{\max}(X)$ and $\max_j u_j(X)$. Each objective is determined by a single server, so either can be attained by several assignments. We therefore define the expected-load optimum as the assignment with the smallest $RR_{\max}$ among those minimizing $\max_j u_j(X)$, and the risk-ratio optimum $X^\star$ as the assignment with the smallest $\max_j u_j(X)$ among those minimizing $RR_{\max}(X)$. Each optimum is thus chosen to be as good as possible on the other criterion. Since enumeration is independent of any heuristic, the structural quantities below are properties of the placement problem itself.

\begin{table}[t]
\centering
\begin{threeparttable}
\caption{Structural results on enumerated small instances.}
\label{tab:exact_structure}
\begin{tabular}{lcc}
\toprule
Quantity & $m=3$ & $m=4$ \\
\midrule
$RR_{\max}(X^\star)/I(X^\star)-1$, mean   & 0.53\%  & 3.57\% \\
$RR_{\max}(X^\star)/I(X^\star)-1$, median & 0.33\%  & 0.69\% \\
$RR_{\max}$ excess of the expected-load optimum & 25.66\% & 29.19\% \\
$\max_j u_j$ excess of $X^\star$ & 22.18\% & 27.58\% \\
Bottleneck overlap & 36.18\% & 22.65\% \\
\bottomrule
\end{tabular}
\begin{tablenotes}[flushleft]\footnotesize
\item $X^\star$ is the risk-ratio optimum and $I(X^\star)$ its proportional ideal from Eq.~\eqref{eq:conditional_ideal}. Excess is relative to the exact minimum of the same quantity. Overlap is the fraction of an instance's risk-feasible assignments in which the server with the largest expected utilization is a risk bottleneck. Entries other than the median are means over instances. For $m=4$, overlap is averaged over the 19 instances that admit a risk-feasible assignment.
\end{tablenotes}
\end{threeparttable}
\end{table}

\noindent\textbf{The risk-ratio optimum is typically close to the proportional ideal.}
Table~\ref{tab:exact_structure} summarizes the structure. The excess of $RR_{\max}(X^\star)$ over the ideal $I(X^\star)$ of Eq.~\eqref{eq:conditional_ideal} has median 0.33\% for $m=3$ and 0.69\% for $m=4$, so the proportional balance of Eq.~\eqref{eq:proportional_condition} is nearly attainable even though tasks are indivisible. The $m=4$ instances are less uniform, with a mean of 3.57\% and a largest value of 27.35\%.

\noindent\textbf{Expected-load balance and risk balance identify different assignments and different critical servers.}
Even with ties resolved in its favor, the expected-load optimum carries an $RR_{\max}$ that exceeds that of the exact risk-ratio optimum by 25.66\% for $m=3$ and 29.19\% for $m=4$ on average. The two criteria also disagree about which server is critical. The server with the largest expected utilization is also a risk bottleneck in only 36.18\% ($m=3$) and 22.65\% ($m=4$) of an instance's risk-feasible assignments on average, so expected load does not reliably identify the server that determines the risk objective. Relative to the expected-load optimum, the risk-ratio optimum raises the maximum expected utilization by 22.18\% for $m=3$ and 27.58\% for $m=4$. The two criteria are therefore different placement objectives, and each optimum performs worse under the other criterion.

\begin{table}[t]
\centering
\caption{Distance from the exact risk-ratio optimum on held-out small instances. Mean gap is the mean over instances of $RR_{\max}/RR_{\max}^\star-1$, where $RR_{\max}^\star$ is the exact minimum. The column ``$\le5\%$'' gives the fraction of instances with a gap of at most 5\%, and ``Feas.'' gives the fraction on which the returned assignment is risk-feasible. One $m=4$ instance admits no risk-feasible assignment.}
\label{tab:exact_algorithm}
\setlength{\tabcolsep}{4pt}
\begin{tabular}{lrrrrrr}
\toprule
& \multicolumn{3}{c}{$m=3$} & \multicolumn{3}{c}{$m=4$} \\
\cmidrule(lr){2-4}\cmidrule(lr){5-7}
Method & Mean gap & $\le5\%$ & Feas. & Mean gap & $\le5\%$ & Feas. \\
\midrule
DG          & 36.68\% & 0\%   & 5\%   & 38.33\% & 0\%   & 10\% \\
RG          & 20.63\% & 10\%  & 50\%  & 26.87\% & 0\%   & 20\% \\
SG-$\sigma$ & 7.57\%  & 40\%  & \textbf{100\%} & 8.45\%  & 30\%  & \textbf{95\%} \\
RA-FF      & 13.73\% & 15\%  & 90\%  & 18.45\% & 20\%  & 55\% \\
Exp-LS      & 29.12\% & 0\%   & 15\%  & 34.78\% & 0\%   & 5\%  \\
RBP         & \textbf{2.64\%} & \textbf{95\%} & \textbf{100\%}
            & \textbf{1.77\%} & \textbf{100\%} & \textbf{95\%} \\
\bottomrule
\end{tabular}
\end{table}

\noindent\textbf{RBP stays close to the exact risk-ratio optimum.}
Table~\ref{tab:exact_algorithm} compares the online methods with the enumerated risk-ratio optimum. RBP has mean gaps of 2.64\% and 1.77\%, and its gap is at most 5\% on 95\% and 100\% of the instances. SG-$\sigma$, the best of the compared greedy rules here, reduces the mean gap from 20.63\% and 26.87\% for RG to 7.57\% and 8.45\%. One $m=4$ instance admits no risk-feasible assignment at all, and RBP returns a risk-feasible assignment on every instance that admits one.

\subsection{Main Performance}
\label{subsec:main_results}

We compare all methods under the default configuration at the two main pressure points.

\noindent\textbf{RBP attains the lowest worst-server overload probability at both pressures.}
Table~\ref{tab:main_results} reports the main multi-period results. At $\rho_R=0.90$ RBP reaches 2.299\% against 3.105\% for SG-$\sigma$, 3.147\% for RG and 3.313\% for RA-FF, which are reductions of 26.0\%, 26.9\% and 30.6\%. At $\rho_R=0.95$ it reaches 2.941\%, reducing the same metric by 18.0\%, 23.9\% and 23.5\% against the same three baselines. All six paired comparisons are statistically significant after Holm correction.

\begin{table*}[t]
\centering
\setlength{\tabcolsep}{4pt}
\small
\caption{Main results over 30 runs. The primary metric is reported as mean $\pm$ standard deviation across runs, and the remaining entries are means. Overload probabilities and risk-feasible rates are in percent.}
\label{tab:main_results}
\begin{tabular}{llccccc}
\toprule
$\rho_R$ & Method & Worst-server overload & Any-server overload & Risk-feasible rate & Total backlog & Max-server backlog \\
\midrule
\multirow{6}{*}{0.90}
& DG          & $4.404\pm0.515$ & 18.327 & 18.83 & 0.004558 & 0.043672 \\
& RG          & $3.147\pm0.574$ & 18.634 & 71.03 & 0.003449 & 0.042614 \\
& SG-$\sigma$ & $3.105\pm0.522$ & 18.426 & 92.13 & 0.004487 & 0.042974 \\
& RA-FF      & $3.313\pm0.590$ & 17.355 & 94.93 & 0.003398 & 0.032776 \\
& Exp-LS      & $4.283\pm0.509$ & 18.188 & 22.73 & 0.004552 & 0.043281 \\
& RBP         & $\mathbf{2.299\pm0.523}$ & \textbf{11.386} & \textbf{96.50} & \textbf{0.002917} & \textbf{0.028122} \\
\midrule
\multirow{6}{*}{0.95}
& DG          & $5.486\pm0.655$ & 22.879 & 4.80 & 0.006035 & 0.056802 \\
& RG          & $3.866\pm0.558$ & 18.834 & 38.47 & 0.004693 & 0.043992 \\
& SG-$\sigma$ & $3.587\pm0.561$ & 23.213 & 65.37 & 0.006056 & 0.056790 \\
& RA-FF      & $3.847\pm0.543$ & 19.354 & 73.57 & 0.004194 & 0.040949 \\
& Exp-LS      & $5.368\pm0.663$ & 22.724 & 5.60 & 0.006003 & 0.056238 \\
& RBP         & $\mathbf{2.941\pm0.557}$ & \textbf{14.044} & \textbf{78.63} & \textbf{0.003744} & \textbf{0.036658} \\
\bottomrule
\end{tabular}
\end{table*}

\noindent\textbf{Accounting for fluctuation has a larger effect than $\sigma$-ordering.}
Two comparisons among the baselines separate the ingredients that RBP also uses. DG, which ignores fluctuation, exceeds RG on the primary metric by 39.97\% and 41.90\% at the two pressures. Adding the $\sigma$-descending order to the same receiver rule, which is the step from RG to SG-$\sigma$, reduces the primary metric by 1.3\% and 7.2\% while raising the risk-feasible rate from 71.03\% to 92.13\% and from 38.47\% to 65.37\%. Since SG-$\sigma$ already uses that ordering, the reductions reported above for RBP over SG-$\sigma$ are obtained through the receiver rule and the refinement of RBP.

Despite a 100~ms budget, four times that of RBP, Exp-LS reaches worst-server overload probabilities of 4.283\% and 5.368\%, only 2.8\% and 2.1\% below DG, against 2.299\% and 2.941\% for RBP. Additional search under the expected-load objective thus does not reproduce the risk reduction of RBP, and that reduction is not explained by search time alone. Section~\ref{subsec:runtime} further shows that the primary metric of RBP changes little between budgets of 5 and 100~ms.

\noindent\textbf{RBP also has the lowest probability that any server overloads.}
The corresponding probabilities are 11.386\% and 14.044\%, which are 34.4\% and 25.4\% below the best compared baseline on this metric at $\rho_R=0.90$ and $0.95$, respectively.

RBP also attains the highest risk-feasible rate at both pressures. At $\rho_R=0.90$ it returns a risk-feasible assignment in 96.50\% of periods against 94.93\% for RA-FF and 92.13\% for SG-$\sigma$, and at $\rho_R=0.95$ the values are 78.63\%, 73.57\% and 65.37\%.

\noindent\textbf{RBP has the lowest backlog at both pressures.}
Relative to RA-FF, the best baseline on both backlog measures, RBP lowers total backlog by 14.2\% and 10.7\% and maximum single-server backlog by 14.2\% and 10.5\%, so its risk reduction is not accompanied by higher backlog in the tested conditions. A lower worst-server overload probability does not by itself imply less backlog, however, since SG-$\sigma$ has a lower worst-server overload probability than RA-FF at both pressures but more backlog than both RA-FF and RG.

\subsection{Ablation Study}
\label{subsec:ablation}

We use two ablations to distinguish what refinement contributes from what the bottleneck-focused visiting order contributes. \emph{Construction only} runs the risk-guided construction of Section~\ref{subsec:construction} and returns its assignment without refinement. \emph{Random-order scan} keeps that construction, the acceptance rule of Eq.~\eqref{eq:acceptance}, the relocation and swap stages and the commit rule of the refinement stage, but visits source servers and their tasks in random order instead of in decreasing order of risk ratio and removal gain.

\begin{table}[t]
\centering
\caption{Ablation results under the main configuration. Risk is the worst-server overload probability, and each comparison is a separate paired experiment.}
\label{tab:ablation}
\begin{tabular}{lccccc}
\toprule
& & \multicolumn{2}{c}{Risk (\%)} & \multicolumn{2}{c}{Feasible rate (\%)} \\
\cmidrule(lr){3-4}\cmidrule(lr){5-6}
Comparison & $\rho_R$ & RBP & Variant & RBP & Variant \\
\midrule
Construction only & 0.90 & 2.298 & 2.426 & 96.17 & 92.70 \\
Construction only & 0.95 & 3.015 & 3.086 & 79.63 & 68.40 \\
Random-order scan & 0.90 & 2.304 & 2.337 & 96.13 & 96.03 \\
Random-order scan & 0.95 & 3.003 & 3.099 & 80.00 & 79.80 \\
\bottomrule
\end{tabular}
\end{table}

\noindent\textbf{Refinement further reduces risk and improves feasibility, with the larger feasibility gain under higher load.}
Table~\ref{tab:ablation} reports both comparisons. Starting from the same constructed assignment, refinement lowers the worst-server overload probability by a further 5.28\% at $\rho_R=0.90$ and 2.30\% at $\rho_R=0.95$, and raises the risk-feasible rate from 92.70\% to 96.17\% and from 68.40\% to 79.63\%, the latter an increase of 11.23 percentage points.

\noindent\textbf{The bottleneck-focused order lowers risk relative to a random order.}
Relative to the random-order scan, RBP lowers the worst-server overload probability by 1.39\% at $\rho_R=0.90$ and 3.09\% at $\rho_R=0.95$, while the feasible rates differ by 0.10 and 0.20 percentage points.

\subsection{Load Sensitivity}
\label{subsec:load_sensitivity}

We vary $\rho_R$ from 0.80 to 0.98 at the default system size to test whether the advantage of RBP persists outside the two main operating points.

\begin{figure*}[t]
    \centering
    \includegraphics[width=\textwidth]{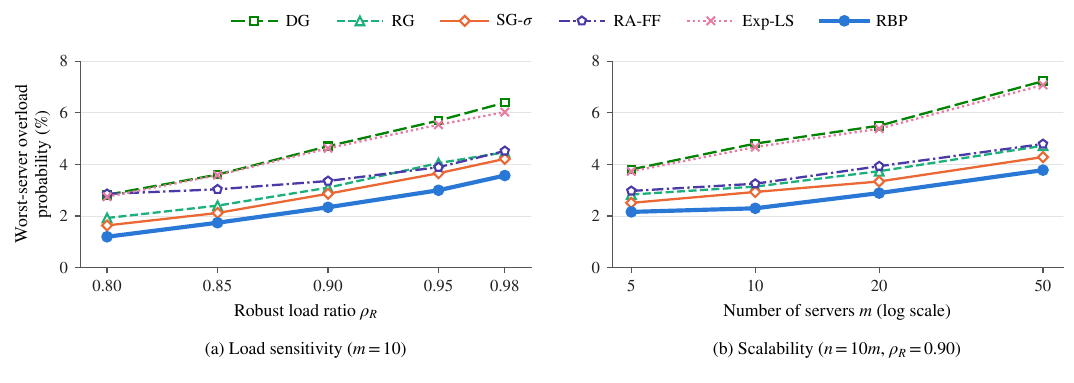}
    \caption{Worst-server overload probability under varying workload pressure and system scale. Each point is the mean over 30 runs. (a) Robust load ratio $\rho_R$ from 0.80 to 0.98 with $m=10$ servers. (b) $m\in\{5,10,20,50\}$ servers with $n=10m$ tasks per period at $\rho_R=0.90$, with capacity recalibrated for each size.}
    \label{fig:load_scale}
\end{figure*}

\noindent\textbf{The advantage persists across the tested pressure range.}
As Fig.~\ref{fig:load_scale}(a) shows, the absolute overload probability rises with pressure for every method. RBP has the lowest worst-server overload probability at all five load levels, and its reduction relative to SG-$\sigma$ ranges from 15.3\% to 26.8\% across the tested range.

Risk feasibility decreases as $\rho_R$ grows. RBP is risk-feasible in 99.80\%, 99.50\%, 95.73\%, 78.80\% and 59.53\% of periods across the five levels. At $\rho_R=0.80$ and 0.85, RBP, SG-$\sigma$ and RA-FF all remain above 99\%, and the differences become larger at the three heavier levels, where RA-FF reaches 52.50\% and SG-$\sigma$ 44.27\% at the highest pressure.

\subsection{Scalability}
\label{subsec:scalability}

We scale the number of servers and tasks proportionally, $m\in\{5,10,20,50\}$ with $n=10m$, to test whether the relative behavior persists as the placement problem grows. Task density per server is held constant and capacity is recalibrated for each configuration by Eq.~\eqref{eq:capacity_calibration}.

\noindent\textbf{The ordering of the methods is preserved as the system grows.}
As Fig.~\ref{fig:load_scale}(b) shows, the absolute worst-server overload probability increases with scale for every method, while RBP retains the lowest value at every tested size, with reductions of 13.9\%, 21.5\%, 13.5\% and 11.8\% relative to SG-$\sigma$.

The risk-feasible rates of RBP across the same four scales are 90.27\%, 96.30\%, 98.63\% and 98.73\%, above SG-$\sigma$ at 86.27\%, 91.53\%, 96.03\% and 98.00\% and RA-FF at 87.70\%, 94.73\%, 98.10\% and 98.57\%. DG shows the opposite trend, with its risk-feasible rate falling from 38.70\% at $m=5$ to 0.07\% at $m=50$.

\subsection{Runtime and Budget Sensitivity}
\label{subsec:runtime}

RBP is designed for a finite online decision budget. We report its per-period running time under the two main conditions, and then vary the budget itself to examine whether additional search time changes the primary metric.

\begin{table}[t]
\centering
\caption{RBP runtime in the main experiment, 3000 periods per condition.}
\label{tab:runtime}
\begin{tabular}{lrrrr}
\toprule
$\rho_R$ & Median (ms) & p99 (ms) & Max (ms) & $>25$~ms \\
\midrule
0.90 & 24.55 & 24.67 & 25.15 & 0.033\% \\
0.95 & 24.57 & 24.67 & 25.04 & 0.033\% \\
\bottomrule
\end{tabular}
\end{table}

As Table~\ref{tab:runtime} shows, the median runtime is 24.55 and 24.57~ms and the 99th percentile is 24.67~ms in both conditions. One of the 3000 periods in each condition exceeds the nominal budget, by at most 0.15~ms, which is consistent with the cooperative deadline of Section~\ref{subsec:budget}. The 25~ms budget is therefore a practical finite online budget rather than a hard real-time upper bound.

At $m=5$ the median runtime is about 6.01~ms, whereas for $m\ge10$ it is about 24.56--24.57~ms. Algorithm~\ref{alg:rbp} stops early only when refinement finds no committable move, so smaller instances can terminate well before the budget, while the larger configurations typically use nearly the full search horizon.

At the two main pressures, the construction stage of RBP has a median runtime of 0.54--0.71~ms per period, comparable to the 0.25--0.76~ms of the four single-pass baselines, so nearly all of the 25~ms budget is spent on refinement. Exp-LS has a median runtime of 104.44--104.89~ms under its soft 100~ms budget.

\begin{table}[t]
\centering
\caption{RBP budget sensitivity: worst-server overload probability (\%). All budgets are run on the same instances, which are drawn separately from those of Table~\ref{tab:main_results}, so absolute values are not directly comparable with Table~\ref{tab:main_results}.}
\label{tab:budget_sensitivity}
\begin{tabular}{rcc}
\toprule
Budget (ms) & $\rho_R=0.90$ & $\rho_R=0.95$ \\
\midrule
5   & 2.306 & 3.164 \\
10  & 2.298 & 3.162 \\
25  & 2.291 & 3.160 \\
50  & 2.320 & 3.145 \\
100 & 2.295 & 3.143 \\
\bottomrule
\end{tabular}
\end{table}

Table~\ref{tab:budget_sensitivity} varies the nominal budget from 5 to 100~ms. Across this range the primary metric varies by 1.27\% of its level at $\rho_R=0.90$ and 0.67\% at $\rho_R=0.95$, so over the tested range the result has limited sensitivity to the online budget. We use 25~ms as the practical default.
\section{Conclusion}
\label{sec:conclusion}

This paper studied online task placement on edge servers when only the mean and variance of each task's workload are known at decision time, a setting in which balancing expected load need not balance overload risk. For independent tasks, placement changes neither the total workload variance nor the total residual capacity of a period, only how fluctuation is paired with headroom across servers. The ratio of a server's fluctuation to its residual capacity captures this pairing and parameterizes the server's worst-case overload probability over all aggregate workload distributions with the given mean and variance. Minimizing the largest ratio relieves the server that forms the risk bottleneck.

RBP uses this structure in two stages that share one finite online budget. It processes tasks in descending order of standard deviation, assigning each to the server with the smallest post-assignment risk ratio, and then refines the resulting assignment through relocations and swaps aimed at reducing the bottleneck risk ratio.

On small instances solved exactly by enumerating every assignment, even the least risky expected-load optimum carries a bottleneck risk ratio 25.66\% and 29.19\% above the exact minimum on average, while the corresponding gaps for RBP are 2.64\% and 1.77\%. The multi-period evaluation uses trace-derived workloads whose task realizations are resampled independently and whose moments are known to the scheduler without estimation error. In this setting, RBP lowers the worst-server overload probability by 26.0\% and 18.0\% relative to the best compared baseline at robust load ratios of 0.90 and 0.95, and keeps the lowest value at every tested load ratio from 0.80 to 0.98 and every tested scale from 5 to 50 servers. An expected-load local search given four times the budget of RBP lowers this probability by less than 3\% relative to deterministic greedy. Under the tested conditions, directly optimizing the bottleneck risk ratio achieves a reduction in worst-server overload probability that additional search under an expected-load objective does not reproduce.

\section*{Funding}
This work was supported by the Inner Mongolia Key Research and Development and Achievement Transformation Program [grant number 2026YFSH0011] and the Inner Mongolia Innovation Capacity Improvement Program [grant number 2026CXHZ0003]. The funders had no role in the study design, the collection, analysis and interpretation of data, the writing of the report or the decision to submit the article for publication.

\section*{CRediT authorship contribution statement}
\textbf{Lei Wang:} Conceptualization, Methodology, Software, Validation, Formal analysis, Investigation, Data curation, Writing -- original draft, Visualization.
\textbf{Winston K.G. Seah:} Supervision, Writing -- review \& editing.
\textbf{Gang Xu:} Resources, Supervision, Project administration, Funding acquisition, Writing -- review \& editing.
\textbf{Xiaodong Xu:} Supervision.
\textbf{Baoqi Huang:} Supervision.

\section*{Declaration of competing interest}
The authors declare that they have no known competing financial interests or personal relationships that could have appeared to influence the work reported in this paper.

\section*{Data availability}
The Google cluster-usage traces (ClusterData 2019) used in this study are publicly available~\cite{google_trace}. The processed task pool and the preprocessing scripts used to construct it are available on Zenodo~\cite{rbp_dataset}.

\section*{Declaration of generative AI and AI-assisted technologies in the manuscript preparation process}
During the preparation of this work, the authors used ChatGPT (OpenAI) and Claude (Anthropic) in order to improve the language and readability of the manuscript. After using these tools, the authors reviewed and edited the content as needed and take full responsibility for the content of the published article.

\bibliographystyle{IEEEtran}
\bibliography{refs}

\end{document}